\documentclass[11pt]{article}
\usepackage[utf8]{inputenc}
\usepackage{amsmath}
\usepackage{amsfonts}
\usepackage{amssymb}
\usepackage{graphicx}
\usepackage{geometry}
\usepackage[numbers,sort&compress,round]{natbib}

\usepackage{url}
\usepackage{subcaption}
\usepackage{booktabs}
\usepackage{siunitx}
\usepackage{fancyhdr}
\usepackage{times}

\usepackage[colorlinks=true,citecolor=black,linkcolor=black,urlcolor=black]{hyperref}
\renewcommand{\thefigure}{\textbf{\arabic{figure}}}

\fancypagestyle{custom}
{      
\fancyhead[L]{}
\fancyhead[C]{}
\fancyhead[R]{}
\fancyfoot[c]{}
\fancyfoot[R]{{\thepage}}
\fancyfoot[L]{\scriptsize peer reviewed manuscript}
}

\begin{document}

\pagestyle{custom}
\begin{center}
\LARGE {\bf Direct Estimation of Earthquake Source Properties from a Single CCTV Camera \\[12pt]}

\normalsize 
Soumaya Latour$^{1\dagger}$, Mathias Lebihain$^{2}$, Harsha S. Bhat$^{3}$,
Cédric Twardzik$^{4}$, Quentin Bletery$^{4}$, Kenneth W. Hudnut$^{5}$,
François Passelègue$^{*4\dagger}$ \\[12pt]
\end{center}

\noindent
{\small
$^{1}$Université de Toulouse, CNRS, Observatoire Midi-Pyrénées, IRAP, Toulouse, France.\\
$^{2}$Navier, ENPC, Institut Polytechnique de Paris, Université Gustave Eiffel, CNRS, Marne-la-Vallée, France.\\
$^{3}$Laboratoire de Géologie, École Normale Supérieure, CNRS, UMR 8538, PSL Université, Paris, France\\
$^{4}$Université Côte d'Azur, CNRS, Observatoire de la Côte d'Azur, IRD, Géoazur, Sophia Antipolis, France.\\
$^{5}$Southern California Edison, Rosemead, 91770, CA, United States.\\
$^{*}$Corresponding author. Email: \href{mailto:francois.passelegue@cnrs.fr}{francois.passelegue@cnrs.fr}\\
$^{\dagger}$These authors contributed equally to this work.}

\begin{abstract}
We present a direct measurement of the slip-rate function from a natural coseismic rupture, recorded on March 28, 2025, during the $M_w$ 7.7 Mandalay earthquake (Myanmar). This measurement was made on video footage of the surface rupture captured by a security camera located only meters away from the fault trace. Using direct image analysis, we measured the relative slip at each time step and deduced the slip rate. Our results show a local slip duration of 1.4 s and cumulative slip of $\sim$3 m, during which surface slip velocity peaked at $\sim$3.5 m/s with passage of the rupture front. These findings demonstrate the pulse-like nature of the seismic rupture, at the location of the recording. Using slip-pulse elastodynamic rupture models, we obtain the complete mechanical properties of this pulse, including the energy release rate.
\end{abstract}

Estimating the spatial and temporal evolution of slip along fault interfaces is critical for understanding the physics of deformation processes in the Earth's crust throughout the seismic cycle \cite{ref1}. Because slip typically occurs at depth under extreme conditions, direct in-situ observations are in most cases impossible. Reconstructions of slip history rely on inverse modeling \cite{ref2}. Consequently, our understanding of earthquake physics is fundamentally limited by the resolution and coverage of the data used in these inversions, as well as the inherent complexity of the forward problem \cite{ref3,ref4,ref5,ref6}. As a result, many key aspects of earthquake rupture, particularly the local dynamics of slip and the associated stress evolution at the fault interface, remain poorly constrained.

On March 28, 2025, a devastating $M_w$ 7.7 earthquake struck Myanmar along the Sagaing fault \cite{ref7} near Mandalay. The earthquake caused catastrophic damage: more than 5,400 fatalities, over 11,000 injuries, and thousands reported missing. Infrastructure losses included 120,000 homes, 2,500 schools, numerous temples, and key transportation networks such as bridges and airports. Several historic sites, particularly in Inwa, suffered extensive damage. This right-lateral strike-slip event ruptured over 450 km of the fault, with slip reaching the surface on long segments, and horizontal displacements up to 6 meters (Figure \textbf{1A} \cite{ref8}). The moment source function derived from the United States Geological Survey (USGS) \cite{ref9} indicates that the rupture propagation lasted approximately 120 seconds (Figure \textbf{1B}), in agreement with the SCARDEC solution \cite{ref10}, and likely included supershear phases \cite{ref11}.

\begin{figure}[t!] 
	\centering
	\includegraphics[width=0.6\textwidth]{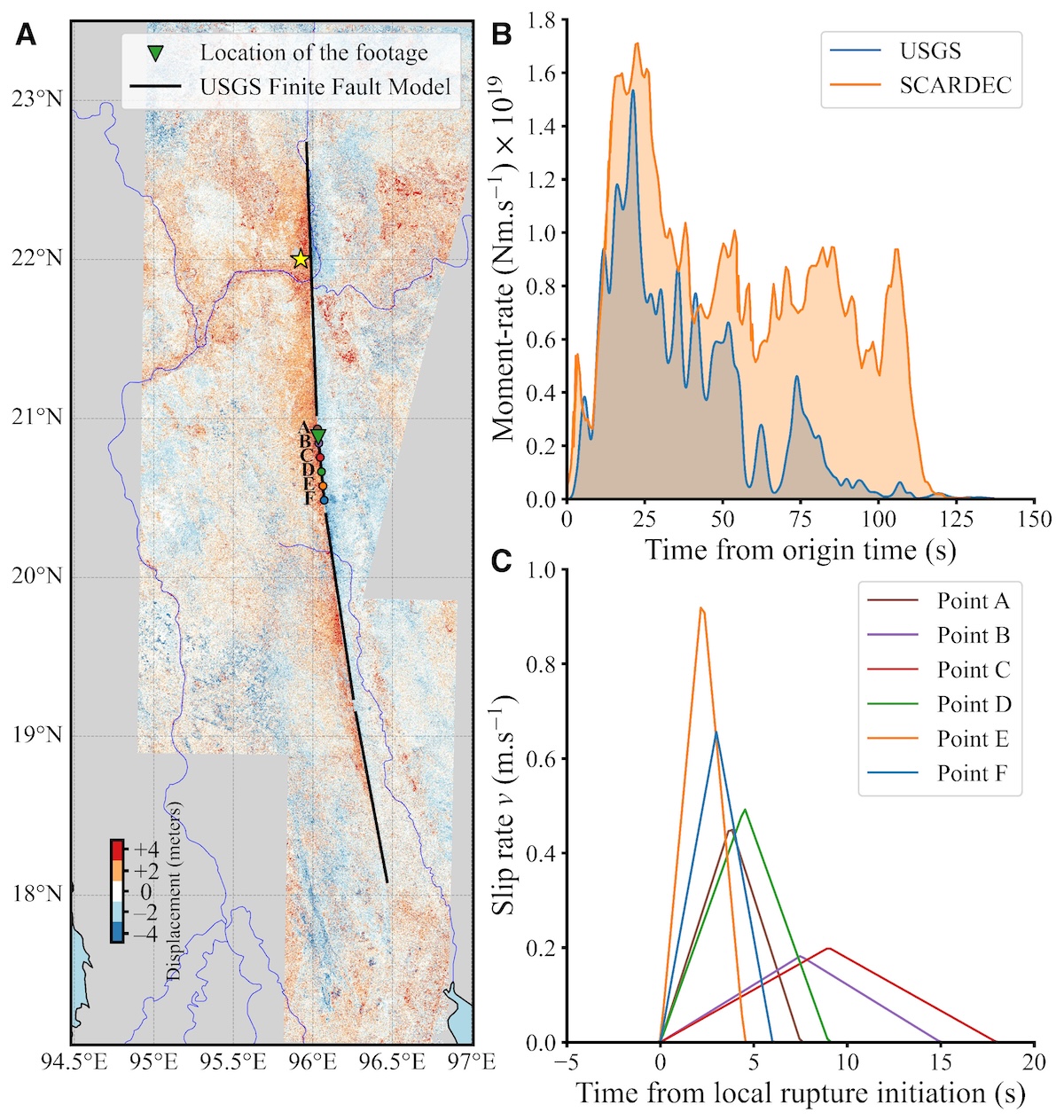} 
	\caption{\textbf{Displacement field, moment-rate functions, and local slip-rate functions for the 2025 Mandalay earthquake.} \textbf{(a)} North-south displacement field derived from the sub-pixel correlation of Sentinel-2 optical imagery. The yellow star indicates the location of the epicentre, while the green triangle marks the location of the footage, as given in the YouTube video description. The black lines show the surface trace of the finite fault model obtained from the USGS. The dots labeled A to F show the locations of the shallowest subfaults of the USGS finite fault model, which were used to extract the local slip-rate functions shown in \textbf{(c)}. \textbf{(b)} Moment-rate functions from the USGS and SCARDEC. \textbf{(c)} Local slip-rate functions derived from the USGS finite fault model. This was only done for the shallowest subfaults of the segment highlighted in \textbf{(a)}. Note that we use a triangular function to display the local slip-rate function instead of the asymmetric cosine function used by the USGS for the kinematic inversion.}
	\label{fig:1} 
\end{figure}

\begin{figure}[t!] 
	\centering
	\includegraphics[width=1\textwidth]{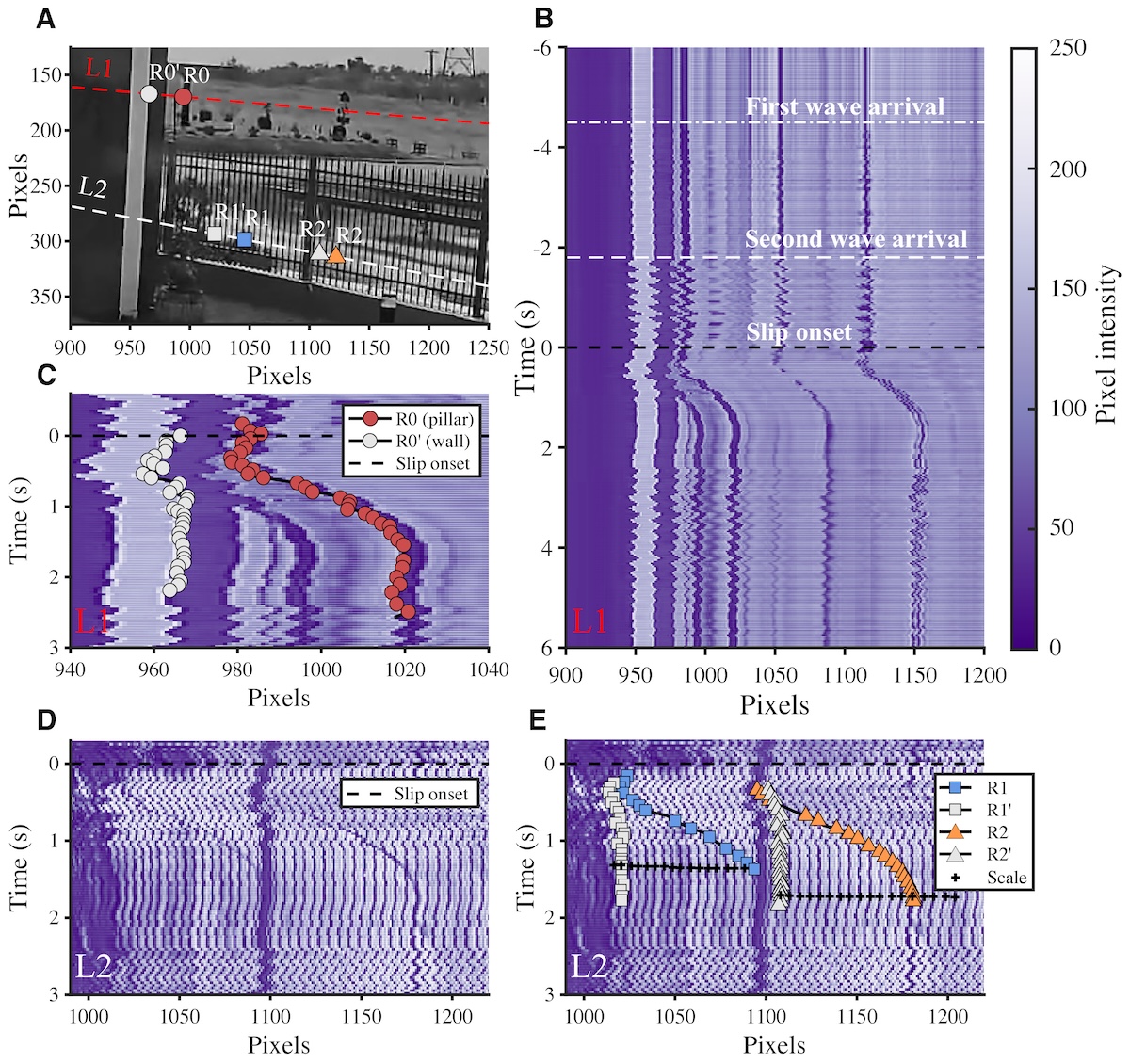} 
	\caption{\textbf{Landmarks tracking method.} 
\textbf{(a)} Zoomed-in video frame showing tracked landmarks and their corresponding reference objects. Landmarks: R0 (red circle) – front-side pillar; R1 (blue square) – pole visible behind the fence; R2 (orange diamond) – second pole behind the fence. Reference points: R0' – arch wall (gray circle); R1', R2' – fence bars near the initial positions of R1 and R2 (gray square and diamond, respectively). The red dashed line (Line 1) indicates the axis used for projections in panels \textbf{(b)} and \textbf{(c)}. The white dashed line (Line 2) indicates the axis used for projections in panels \textbf{(d)} and \textbf{(e)}. 
\textbf{(b)} Motion tracking with annotations showing the arrival of the first seismic wave, second wave, and slip onset. \textbf{(c)} Pixel displacement time series for R0 and R0', measured along Line 1. \textbf{(d)} Pixel displacement time series for R1, R1', R2, and R2' along Line 2. \textbf{(e)} Same as \textbf{(d)}, with additional annotations showing the tracking of each landmark and associated reference for R1 and R2, and a motion scale (black crosses)
}
	\label{fig:2} 
\end{figure}

Beyond its societal impact, the earthquake offered an unprecedented scientific opportunity: for the first time, a CCTV security camera located just meters from the surface rupture captured the real-time deformation of the ground during seismic faulting \cite{ref12}. This footage provides a unique and direct observation of coseismic surface displacement, enabling us to extract the time history of slip and slip-rate at a specific fault point, measurements that, until now, were only accessible through laboratory experiments or inferred indirectly via modeling. In this study, we analyze the footage to obtain a direct measurement of the slip-rate function during a natural earthquake \cite{ref13}, and additionally, to invert the evolution of shear stress, energy dissipation (breakdown work), and rupture dynamics. Our results provide ground-truth constraints on seismic rupture physics, offering a critical benchmark for validating numerical models and seismic inversions, and helping to close a longstanding observational gap in earthquake science.

\section*{Measurement of local slip rate during the earthquake}

The camera was positioned on the east side of the fault and was approximately oriented in the southwest direction. To estimate the local slip rate, we tracked distinct visual landmarks located on both sides of the fault (Figure \textbf{2A}). The primary reference point was a concrete or metallic pillar ($R_0$) located on the west side of the fault, while two additional landmarks $R_1$ and $R_2$ were small poles located adjacent to a road and behind a fence composed of vertical bars. Throughout the video, these features display predominantly horizontal motion relative to the foreground, consistent with right-lateral strike-slip faulting.

\begin{figure}[t!] 
	\centering
	\includegraphics[width=\textwidth]{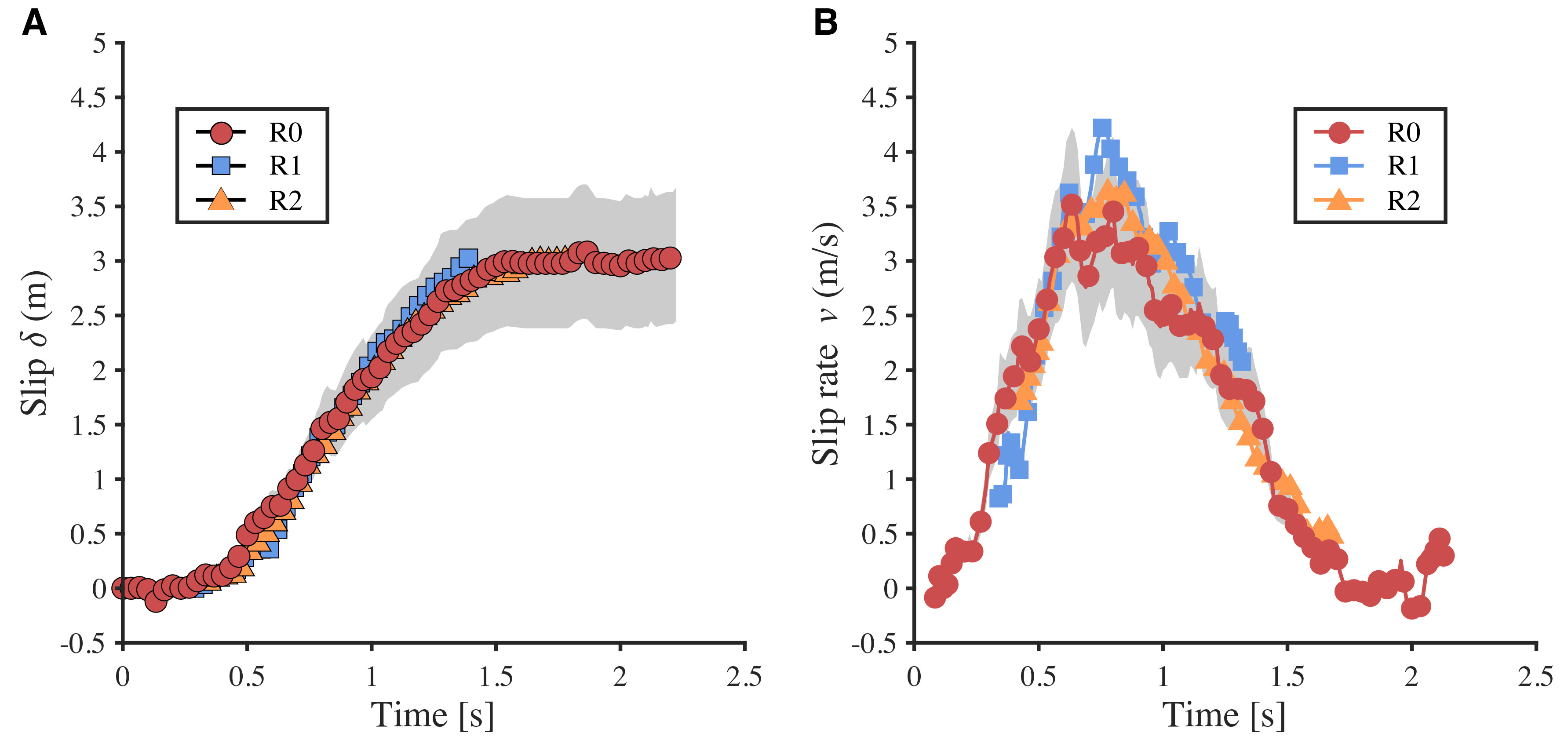} 
	\caption{\textbf{Slip and slip rate functions measured at the camera location (a)} Evolution of the fault slip during the rupture propagation modeled using the three different reference points (Figure \ref{fig:2}a). \textbf{(b)} Slip-rate function derived from (a) for the three different references. The gray shaded areas in \textbf{(a)} and \textbf{(b)} represent the effect of the 20\% uncertainty in our slip estimate on the measurements derived from $R0$, which is the reference used in the modeling. The symbols correspond to those described in Figure \ref{fig:2}.}
	\label{fig:3} 
\end{figure}

To quantify this motion, we defined two fault-parallel lines intersecting the vertical landmarks (Figure \textbf{2A}) and tracked the pixel intensity along these lines over the entire duration of observable rupture propagation (Figure \textbf{2B-E}). We defined $t = 0$ s as the approximate onset of fault slip (corresponding to the time stamp 12:46:34.633 of the video and not synchronised to UTC). Ground shaking began at $t = -4.5$ s, evident as coherent motion across the frame due to camera motion. A secondary seismic phase was observed at approximately $t = -1.8$ s. The onset of fault slip was identified as the moment when the tracked landmarks began to move relative to nearby, foreground features. This relative motion continued until approximately $t = 1.6$ s (Figure \textbf{2C}).

For each landmark, we selected a nearby visually stable reference point on the east side of the fault, as shown in Figure \textbf{2A}: $R_0'$ (a vertical wall), and $R_1'$ and $R_2'$ (vertical bars of the fence). We tracked the motion of both the landmark and its corresponding reference feature over time (Figure \textbf{2C,E}). The difference in their pixel positions provided a measure of relative displacement that was corrected for camera motion. The temporal and spatial resolutions were determined using the method described in the supplementary material (Figure \textbf{S1}), and a spatial scaling was applied using an estimate of the final slip. This approach yielded a temporal resolution of 0.033 s and a spatial resolution with a relative uncertainty of approximately $\pm 20\%$.

The slip history derived from the three measurements is shown in Figure \textbf{3A}. The curve obtained from $R_0$ provides the most reliable estimate, as both the onset and arrest of motion were clearly visible in the video. The timing and slope of estimates from $R_1$ and $R_2$ are consistent with that of $R_0$. The slip increased smoothly from 0 to 3 m over approximately 1.4 s, after which it ceased. The slip-rate function (Figure \textbf{3B}) was obtained by numerically differentiating the slip history after resampling it at 90 Hz. A moving average with a 0.33 s time window was then applied, as we interpreted the oscillations in the raw curve to result from uncertainties in manual tracking and camera-motion correction. The slip-rate function is slightly asymmetric: the slip rate rose rapidly from 0 to 3.5 m/s within 0.6 s, then decreased more gradually to 0 over approximately 1 s. These values correspond to an average slip acceleration of 5.8 m/s$^2$ and a slip deceleration of 3.5 m/s$^2$.

The total slip duration is best estimated using the rise time of the slip function, as the smoothing applied during signal processing tends to artificially broaden the slip-rate pulse. The most robust estimate of slip duration is $\Delta t = 1.4$ s (Figure \textbf{3A}), independent of any assumptions related to spatial scaling. In contrast, our estimates of total and intermediate slip must consider a relative uncertainty of approximately $\pm 20\%$ in the spatial scaling factor. The uncertainty in total slip also affects absolute values of the slip rate.

\section*{Duration and shape of the slip rate function}

Our results highlight the much shorter local slip duration (1.4 s) compared to the total rupture duration of 100-120 s (Figure \textbf{1B}). This clearly shows that the rupture had a pulse-like nature, also called self-healing pulse, at least at the surface along this segment of the fault. Self-healing pulses have long been identified as a possible mode of earthquake rupture \cite{ref14}. In fact, pulse-like behavior appears to be a fairly common feature of rupture models for large earthquakes \cite{ref15,ref16}. The origin of this pulse-like behavior can be attributed to (i) the elongated geometry of the rupture or (ii) the structure of the fault damage zone \cite{ref17}. In the case of the Mandalay earthquake, the rupture was expected to have quickly saturated the seismogenic width, resulting in a large length-to-width ratio at the footage location. Under these conditions, it is expected that a pulse-like rupture would emerge naturally \cite{ref18,ref19}. However, alternative slip-healing mechanisms may also account for the short 1.4 s slip duration observed here. In particular, thermal pressurization of pore fluids within a thin gouge layer can induce rapid re-strengthening behind the rupture tip. Using representative values for hydrothermal diffusivity ($\alpha = 10^{-4}$ to $10^{-6}$ m$^{-2}$/s) and a gouge thickness $h$ ranging from $10^{-2}$ to $10^{-3}$ m \cite{ref20,ref21}, the expected slip duration scales to $\Delta t \approx h^2/\alpha \approx$ 1 s \cite{ref22}, consistent with our observations. This mechanism does not rely on rupture aspect ratio and could explain the relatively small pulse width compared to the total rupture depth. Our results, however, are agnostic to various healing mechanisms simply due to the lack of additional in situ physical measurements.

The slip-rate values are consistent with coseismic slip rates expected during large earthquakes. Particle velocity in the vicinity of the fault was predicted to be capable of reaching several meters per second for self-healing pulse ruptures \cite{ref14}. Note that our direct measurements of the fault slip rate are much higher than those estimated from inverted teleseismic data \cite{ref9}, which indicated a peak slip rate on the order of 0.2 m/s at the footage location (Figure \textbf{1C}). In addition, the recorded duration of local slip is much shorter than the inverted estimate: 1.4 s against $\approx$ 16 s for the USGS kinematic model. However, it is well known that modeling the rupture process involves significant uncertainty \cite{ref23,ref6}. In fact, the inverted slip duration just 10 km north of the footage location is much shorter $\approx$ 6 s, and the peak slip rate is higher (0.45 m/s) (Figure \textbf{1C}). This discrepancy underscores the value of direct observation of slip during a natural earthquake as an empirical benchmark for validating and refining seismic source models of large earthquakes.

The shape of the slip-rate function is a key parameter in kinematic slip models \cite{ref24} and plays an important role in ground motion prediction \cite{ref25}. In this study, the smoothed slip-rate function exhibits a simple form that can be described as an asymmetric triangle, with the duration of the acceleration phase approximately 60\% that of the deceleration phase. A similarly smooth, asymmetric and triangular slip-rate function has been observed in friction experiments involving self-healing slip pulses propagating at 0.76 times the Rayleigh wave speed \cite{ref26}.

An analytical form of the slip-rate function was proposed for mode III self-healing pulses \cite{ref27}. This form, sometimes referred to as the Yoffe function, is characterized by an infinitely steep onset, a sharply peaked maximum, and a long tail with a variable slope. Although the Yoffe function qualitatively captures the asymmetry observed in our data, its overall shape is more peaked and strongly asymmetric. A regularized version of the Yoffe function, derived from dynamic rupture models \cite{ref28}, more closely resembles the slip-rate function observed in our measurements, though it still exhibits greater asymmetry.

\section*{Estimate of dynamic source parameters from the fault slip rate}

\begin{figure}[t!] 
	\centering
	\includegraphics[width=\textwidth]{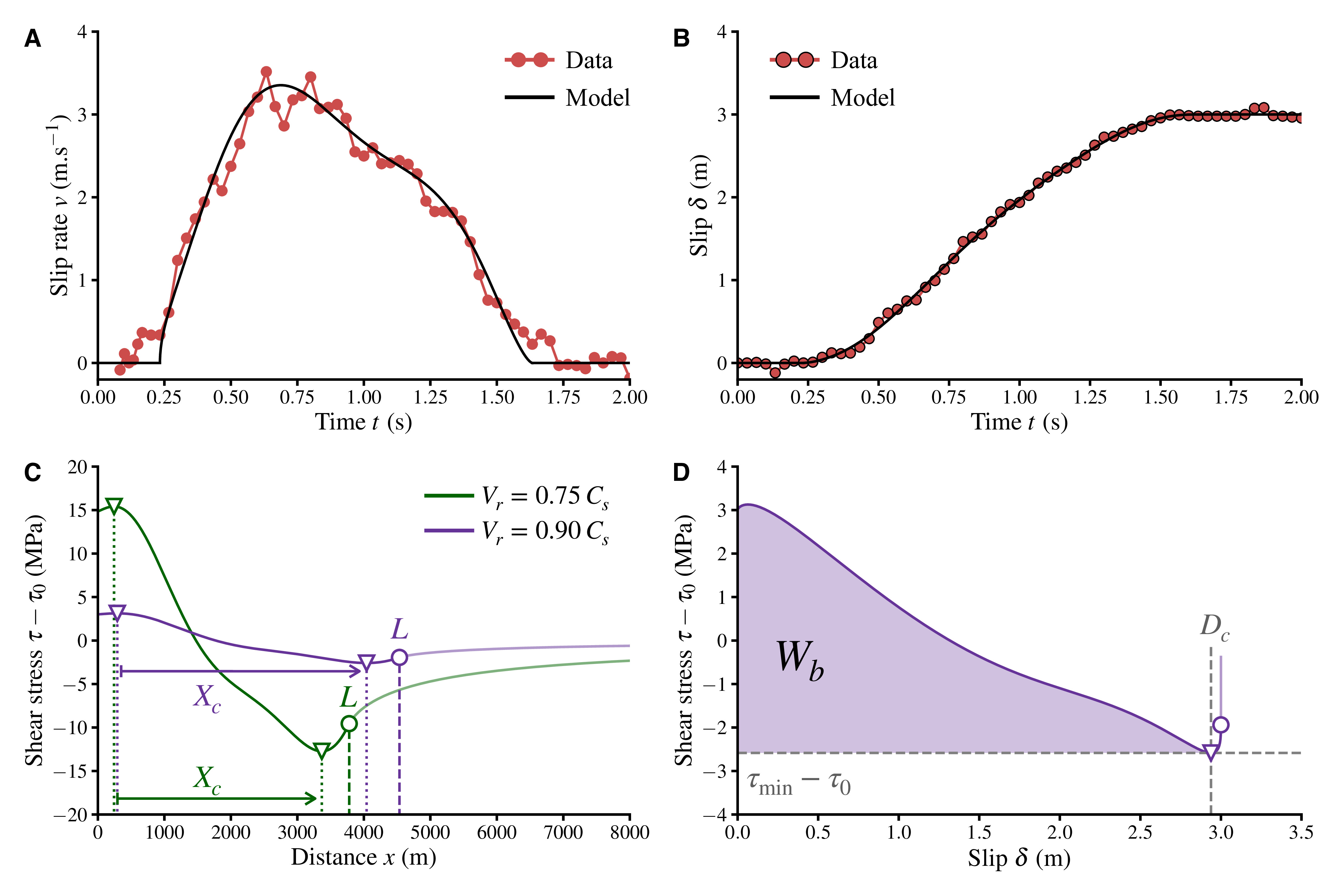} 
	\caption{\textbf{Stress evolution and stress-slip relationship inferred from the slip-rate function}. Direct measurement of (\textbf{a}) fault slip rate and (\textbf{b}) slip (in red circle markers) and Chebyshev-based fit (in black solid line). (\textbf{c}) Solution of the evolution of the shear stress change during the slip pulse propagation for both rupture scenario $V_r = 0.75\,C_s$ (in solid green line) and $V_r = 0.90\,C_s$ (in solid purple line). The cohesive zone size $X_c$ is defined as the distance between the maximum and minimum of the shear stress change, while the pulse length $L$ corresponds to the extent of the actively sliding region. (\textbf{d}) Evolution of stress variation $\tau - \tau_0$ with slip, including the healing stage for $V_r = 0.90\,C_s$. The breakdown work $W_b$ (in shaded purple area) is defined from an integral of the shear stress change $\tau - \tau_\mathrm{min}$ up to the critical slip distance $\delta = D_c$.}
	\label{fig:4} 
\end{figure}

We applied the elastodynamic equilibrium condition to infer the evolution of shear stress from a slip-based model. Assuming a 2D plane strain shear (mode II) rupture propagating at a constant sub-Rayleigh rupture velocity $V_r$ along the fault, the shear stress $\tau(x)$ is related to the spatial distribution of slip rate $v(x)$ through the singular integral equation \cite{ref22,ref29,ref30}:

\begin{equation}
\tau(x) = \tau_0 - \frac{\bar{\mu}}{2\pi V_r} \text{PV} \int_0^L \frac{v(\xi)}{x - \xi} d\xi
\end{equation}

where $\tau_0$ is the background stress, $\bar{\mu}$ is a scaled shear modulus, which depends on the rupture velocity $V_r$, $x$ is the distance to the rupture tip, $L = V_r \Delta t$ is the length of the slip pulse, which defines the spatial extent of the actively slipping region, and PV denotes the Cauchy principal value of the integral.

To compute the shear stress evolution $\tau(x) - \tau_0$ along the fault, we interpolated $v(x)$ using a Chebyshev-based fit (Figure \textbf{4A}), which captures the asymptotic behavior of slip and slip rate expected from fault mechanics, at both the leading and trailing edges of the pulse (see Figures \textbf{S4}, \textbf{S5} and supplementary material for more details).

Once $v(x)$ was known, we inferred the shear stress evolution from Eq. (1) (Figure \textbf{4C}). Since the exact value of $V_r$ was not known at this stage, we present results for two representative cases in Figure \textbf{4B}: $\frac{V_r}{C_s} = 0.75$ and $\frac{V_r}{C_s} = 0.90$. These ratios correspond to the two end-member rupture velocities of a rupture propagation scenario accounting for the ground motion and arrival times of the first and second seismic-wave arrivals as observed in the footage (Figure \textbf{2B}), implying a local subshear rupture velocity (see supplementary material and Figure \textbf{S3}).

As anticipated for pulse-like ruptures, we observed that the stress drop was followed by a restrengthening phase (Figure \textbf{4C}), which spanned approximately 17\% of the pulse length. The dependence of the shear stress evolution on the rupture velocity $V_r$ was also expected: a lower rupture velocity $V_r$ requires a larger strength drop $\Delta\tau$, defined as the difference between the maximum and the minimum shear stress, to match the measured cumulative slip and slip rate. In addition, the size of the cohesive zone $X_c$, which corresponds to the distance over which the frictional strength of a fault degrades from its peak to residual level, increases linearly with $V_r$. Specifically, we found $\Delta\tau = 5.7$ MPa and $X_c = 3749$ m for $V_r = 0.90C_s$, compared to $\Delta\tau = 28.2$ MPa and $X_c = 3124$ m for $V_r = 0.75C_s$. Examining the apparent stress-slip relationship in Figure \textbf{4C}, we observed a linear decay of shear stress $\tau$ with slip $\delta$, up to a critical slip value $D_c = 2.94$ m, which is consistent for both rupture velocities.

We then calculated the breakdown work $W_b = \int_0^{D_c} (\tau(\delta) - \tau_{\min}) d\delta$ associated with the slip pulse, integrating up to the point of minimum stress $\tau_{\min}$, defined spatially at $X_c$ and in cumulative slip at $D_c$ (Figures \textbf{4B--C}). Because this estimate is strongly dependent on final slip $D_{\text{fin}}$ and rupture velocity $V_r$, we conducted a sensitivity analysis to characterize the uncertainty on these parameters (Figure \textbf{S6}, supplementary material). For the assumed final slip of 3.0 m, $W_b$ ranges from 7.7 MJ/m$^2$ for $V_r = 0.90C_s$ to 38.2 MJ/m$^2$ for $V_r = 0.75C_s$.

To better constrain $V_r$ and provide an independent direct measurement of the energy release rate $G$ during the rupture pulse of the $M_w$ 7.7 Mandalay earthquake, we fitted a classical two-dimensional steady-state rupture pulse model \cite{ref29} to the observed slip pulse. The model assumes a steady-state sub-Rayleigh rupture velocity $V_r$, and a constant-width cohesive zone in which the friction linearly decreases from the peak to the residual frictional strength ($f_p = 0.6$ and $f_r = 0.1$, respectively) behind the rupture front (Figure \textbf{S7a}). Unlike the direct slip-based model used previously, this model imposes a cohesive zone friction law and no healing within the pulse. Therefore, the energy released during the weakening contributes entirely to the rupture propagation and corresponds to $G$ as described in fracture mechanics. While the direct slip-based approach is applicable in this study due to the unique availability of true on-fault slip measurements from CCTV footage, the linear cohesive zone model is a more general framework that can be applied to conventional seismic observations, including data recorded at stations located away from the fault \cite{ref31}. We assumed a nominal normal traction of $\sigma_0 = 10$ MPa, consistent with the shallow nature of the rupture. The slip pulse duration was fixed at $\Delta t = 1.4$ s, as estimated previously. We performed a grid search on the scaled cohesive zone size $\frac{X_c}{L}$ and scaled rupture velocities $\frac{V_r}{C_s}$, identifying the parameters that produce the largest Signal Similarity Index (SSI), which is an Euclidean-norm measure of how well the predicted slip-rate evolution matched that measured from image analysis. As seen in Figure \textbf{5A}, the optimal SSI corresponds $\frac{X_c}{L} = 0.71$ and $\frac{V_r}{C_s} = 0.903$, which provides the fit of Figure \textbf{5B}. This $\frac{X_c}{L}$ value is lower than the elastodynamic estimate of 0.83, as the pulse model prohibits restrengthening within the pulse, reducing the effective weakening (cohesive) zone. Nevertheless, the result supports $V_r = 0.90C_s = 3240$ m/s as the most consistent rupture scenario.

\begin{figure}[t!] 
	\centering
	\includegraphics[width=\textwidth]{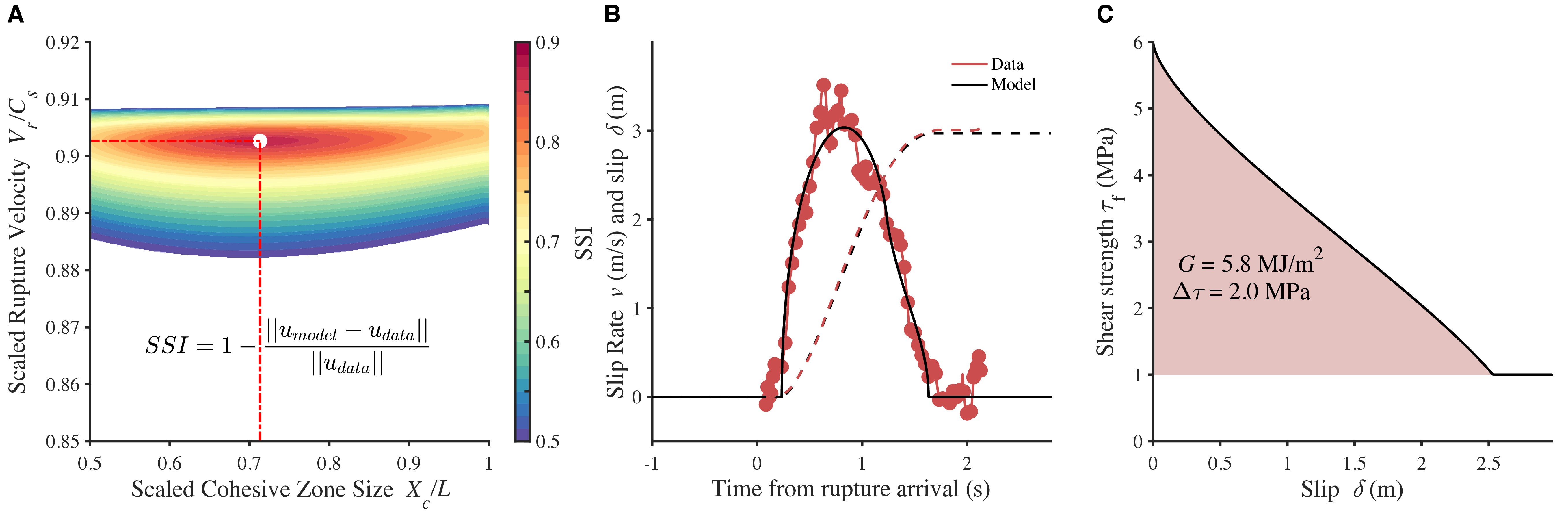} 
	\caption{\textbf{Evaluation of rupture velocity and energy release rate.} (\textbf{a}) Grid search for the best fitting scaled cohesive zone size, $X_c/L$, and scaled rupture velocity, $V_r/C_s$. The fit is measured using the Scaled Similarity Index defined above (\textbf{b}) Comparison, of slip and slip rate, between the best fitting model slip pulse and the observed one  (\textbf{c}) Evolution of friction strength, $\tau_f$, with slip, $\delta$.}
	\label{fig:5} 
\end{figure}

We have now obtained a complete mechanical characterization of the slip pulse observed in the CCTV data. The inferred strength drop was $\tau_p - \tau_r = 5$ MPa, the stress drop was $\tau_0 - \tau_r = 2$ MPa, and the corresponding slip-weakening distance was $X_c \approx 2.5$ m. These values are consistent with the slip-based model estimates ($\tau_p - \tau_r = 5.7$ MPa, $\tau_0 - \tau_r = 2.6$ MPa), except for $D_c = 2.94$ m, which was slightly overestimated due to the larger cohesive zone size. Ultimately, the energy release rate $G$ was computed as $\int_0^{D_c} [\tau_f(\delta) - \tau_r] d\delta = 5.8$ MJ/m$^2$, where $\tau_f(\delta)$ is the strength evolution with slip $\delta$ (Figure \textbf{5C}). Assuming that $G = G_c$, our analysis provides here a direct estimate of the local fracture energy $G_c$ of a natural fault. The energy release rate is of the same order of magnitude as our elastodynamic estimate of the breakdown work $W_b = 7.7$ MJ/m$^2$. While related, the two quantities are not equivalent: $G_c$ is expected to be smaller than $W_b$ for ruptures that have undergone complete frictional weakening \cite{ref32,ref33,ref34}. Our value of $G_c$ is consistent with expectations for an earthquake of the size and geometry of the Mandalay event \cite{ref35,ref36,ref37,ref38}, but it is significantly larger than values typically reported for experimental faults, both in analog materials \cite{ref39,ref40} and in crustal rock samples \cite{ref41,ref42}. This difference is primarily attributed to the larger slip-weakening distance $D_c$ inferred here, compared to the much smaller values typically observed during laboratory earthquakes \cite{ref40,ref41}. Nevertheless, such large values of $D_c$ remain compatible with those measured during high velocity friction experiments under upper-surface stress conditions \cite{ref43,ref44,ref45} and are expected from seismic kinematics inversion of large earthquakes \cite{ref46}. Beyond the specific case of the Mandalay earthquake, these direct measurements offer a rare, empirical benchmark for earthquake source modeling. Unlike traditional kinematic and dynamic inversions, which are limited by data coverage and modeling assumptions, the video-derived slip and slip-rate functions provide ground-truth constraints on the physical processes behind large earthquakes, like fault weakening, energy dissipation, and rupture dynamics.

Finally, our study highlights the potential for a new observational approach in earthquake science: strategically deploying CCTV or high-frame-rate cameras near active, shallow-fault zones. Such installations could capture not only rapid, coseismic fault slip, as demonstrated here, but also slower aseismic deformation or post-seismic creep. Compared to traditional instrumentation such as broadband seismological stations or continuous GPS networks, video monitoring systems are relatively low-cost, widely available, and easy to deploy, making them an attractive complementary tool for expanding fault monitoring capabilities. With appropriate placement and calibration, these instruments could provide direct, high-resolution constraints on fault behavior across the full spectrum of slip modes, opening new avenues for studying earthquake physics.

\section*{Acknowledgments}

The authors are grateful to the persons who uploaded and shared the analyzed footage on social media. F.P. thanks Barnaby Fryer, Nathalie Calza and Federica Paglialunga for discussions and for proofreading the manuscript. S.L. and F.P. thanks Jean-Paul Ampuero and Françoise Courboulex for valuable discussions. Authors thank Dmitry Garagash, an anonymous reviewer and the editor for their constructive remarks, which helped to enhance this paper.

\section*{Funding}

F.P. acknowledge support from the European Union (ERC Starting Grant HOPE num. 101041966). H.S.B. acknowledge support from the ERC Consolidator Grant (865411), PERSISMO, for partial support of this work. Q.B. acknowledges support from the ERC Starting Grant (949221), EARLI, for partial support of this work.

\section*{Author contributions}

S.L. and F.P. conceived the study., S.L. and K.W.H. conducted the frame analysis based on satellite imaging, M.L. developed the elastodynamic inversion, F.P., H.S.B. and M.L. estimated the rupture properties, C.T. and Q.B. analyzed the kinematic inversion from USGS and provided estimates of the wave speeds, S.L. and F.P. wrote the first draft of the manuscript, K.W.H. provided critical data for spatial scaling and validation. All authors contributed to discussions, interpretation of results, and reviewed the manuscript.

\section*{Data and materials availability}

All data and code necessary to directly reproduce our results are available online at \href{https://doi.org/10.5281/zenodo.16785672}{10.5281/zenodo.16785672}.


\setcounter{section}{0}
\renewcommand\thesection{}
\renewcommand\thesubsection{\Alph{section}.\arabic{subsection}}

\renewcommand{\thefigure}{A\arabic{figure}}
\renewcommand{\thetable}{A\arabic{table}}
\renewcommand{\theequation}{A\arabic{equation}}
\setcounter{figure}{0}
\setcounter{table}{0}
\setcounter{equation}{0}
\setcounter{page}{1} 


\section{Appendix}

\subsection{Spatiotemporal scaling of the video of the seismic rupture }

The time scale is accurately given by the video frame rate of 30 frames per second which we verified using the timestamps embedded in the footage, and which is a standard frame rate for a commercial camera.

The use of three landmarks allows us to check for the global shape of the slip function. R0 is the best measure, because its motion is clearly tracked from before the slip onset to after its end. In contrast, the end of the motion of R1, and the beginning of the motion of R2 are missing, because the poles are hidden at these times. However, measurements of R1 and R2 are interesting because they move just a few meter behind the equally spaced bars of the fence. Moreover, the fence and the road along which the poles are installed are almost parallel to the fault (see Figure \ref{fig:SP3}A). Hence, the fence provides a graduated line in the direction parallel to the fault. We could verify that there is negligible distortion in the span of each pole motion. Moreover, we could use the bar spacing to put the motions of R1 and R2 at the same scale. Finally, to adjust for the missing motion of R2 at the beginning, we shifted its displacement value so that the first value, measured at $t = 0.34$~s, equals the value measured at the same time for R1.

We then scaled the displacement of R0 with the final slip value, estimated in the following way. 

\begin{figure} 
	\centering
	\includegraphics[width=1\textwidth]{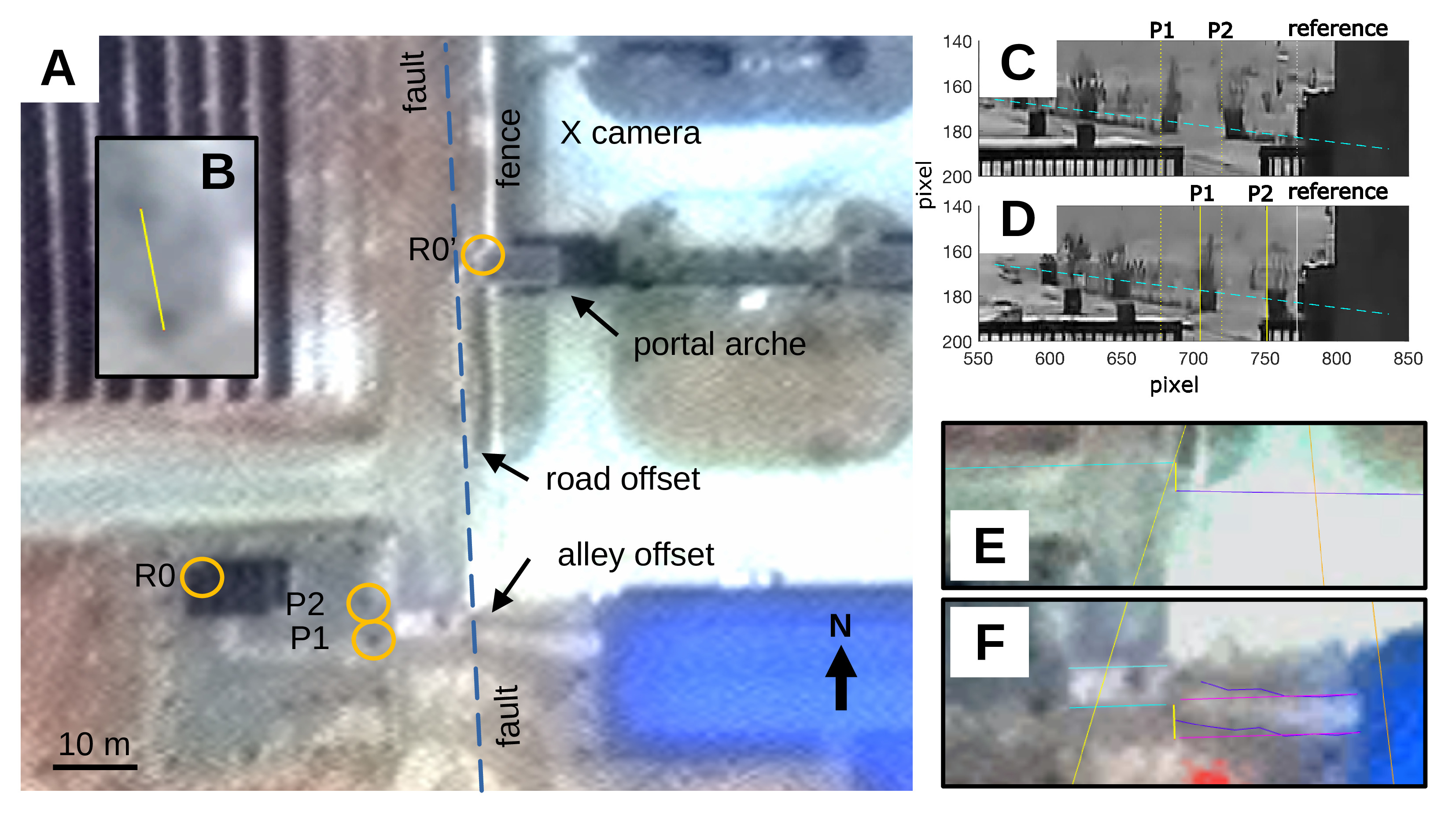}
	\caption{\textbf{Satellite image and scaling process.} \textbf{(A)} Post-rupture satellite image (Maxar-30cm © 2025 Maxar Technologies, original resolution $0.49$~m/pixel, captured at UTC 2025-04-01T03:58:50), showing the approximate locations of key video elements (R0: pillar, R0': arch wall, P1 and P2: plant boxes). The blue dashed line indicates the approximate fault trace, derived from both the satellite image and the video. \textbf{(B)} Zoomed view of P1 and P2. The yellow segment represents $4.67$~m. \textbf{(C–D)} Before-and-after images of P1 and P2, used for scaling. The blue dashed line is a projection line indicating that the two boxes moved parallel to the fault. Yellow lines mark the positions of P1 and P2 along this line. The white line corresponds to a foreground element, used to correct for camera shake. \textbf{(E)} Zoom on a road offset, estimated at $2.8 \pm 0.4$~m (length of yellow segment) using a reference line method \cite{ref54}. \textbf{(F)} Zoom on an alley offset, estimated at $2.9 \pm 0.3$~m (length of yellow segment). Includes content sourced via SkyWatch Space Applications Inc.}
	\label{fig:SP1} 
\end{figure}

Thanks to satellite imagery, the distance between two plant boxes (P1 and P2) can be estimated at approximately $4.67$~m (see Figures \ref{fig:SP1}A and \ref{fig:SP1}B). These boxes are located on the west side of the fault and are almost perfectly aligned in a direction parallel to the fault. We first traced a line, in the video frames, that intersects the two boxes (Figure~\ref{fig:SP1}C), and corresponds roughly to a horizontal line parallel to the fault. We then identified the positions of the boxes along this line just before the onset of slip and again 1.65~s later, when the final slip recorded during the video is reached. The distance along this line between the two boxes serves as a reference scale.

The total displacement is estimated by measuring the difference in position of the southern box, P1, along the line before and after the slip event (see Figure~\ref{fig:SP1}C). The final position of this box lies between the initial positions of the two boxes, which helps to ensure the accuracy of the scaling in this region of the image. This approach yields a total slip of approximately $3.0$~m, with an estimated uncertainty of about $20$~\%, due to uncorrected perspective effects, camera distortion, and the limited resolution and clarity of both the satellite imagery and the original video footage. This estimate is consistent with a tentative slip measurement based on the offset of a linear feature observed in post-rupture satellite imagery (see Figures \ref{fig:SP1}E and \ref{fig:SP1}F).

Finally, we scaled the R0 and the R1-R2 slip curves so that their final slip values reached $3$~m. 

\subsection{Estimation of the medium properties}

\begin{figure}
    \centering
    \includegraphics[width=1.0\textwidth]{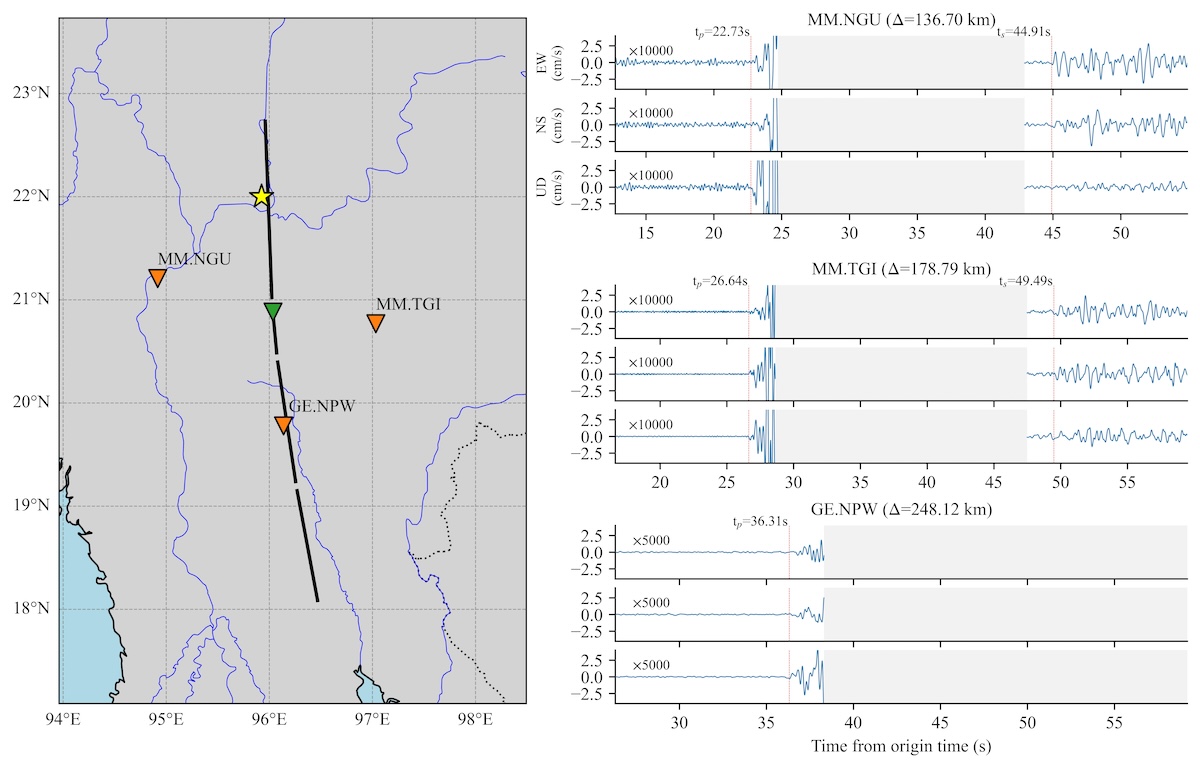}
    \caption{(Left) Map showing seismological stations (orange triangles) located within 300 km of the epicentre (yellow star). The green triangle indicates the location of the CCTV footage site. The black lines show the surface trace of the finite fault model from the USGS. (Right) Three-component velocity waveforms for each station. The red lines indicate the manually picked first arrivals of the P and S waves. The P wave has been enhanced for the three stations to highlight the first arrival, while the S wave remains at scale. The red lines show the manually picked P and S wave first arrivals. We hide in gray the irrelevant portion of the waveforms to ensure better visibility of the first arrivals.}
    \label{fig:SP2}
\end{figure}

To estimate the values of $C_p$ and $C_s$, we used waveforms from three seismological stations of the Myanmar National Seismic Network and the GEOFON Seismic Network \cite{ref47,ref48}, which are less than $300$~km from the epicenter (see Figure \ref{fig:SP2}). We manually pick the first arrivals of the P and S waves at stations NGU and TGI. For station NPW, only the first arrival of the P wave is manually picked due to the possibility of supershear rupture in this part of the fault.

Given the epicenter-receiver distances and the measured first arrival times shown in Figure \ref{fig:SP2}, we calculate the P-wave velocity of the medium ranges between $6000$~m/s and $6800$~m/s. For the S-wave velocities, we obtained values ranging from $3000$~m/s to $3600$~m/s. In the following we use $C_s=3600$~m/s and $C_p=6235$~m/s.

\subsection{Local rupture speed}

\begin{figure} 
	\centering
	\includegraphics[width=0.9\textwidth]{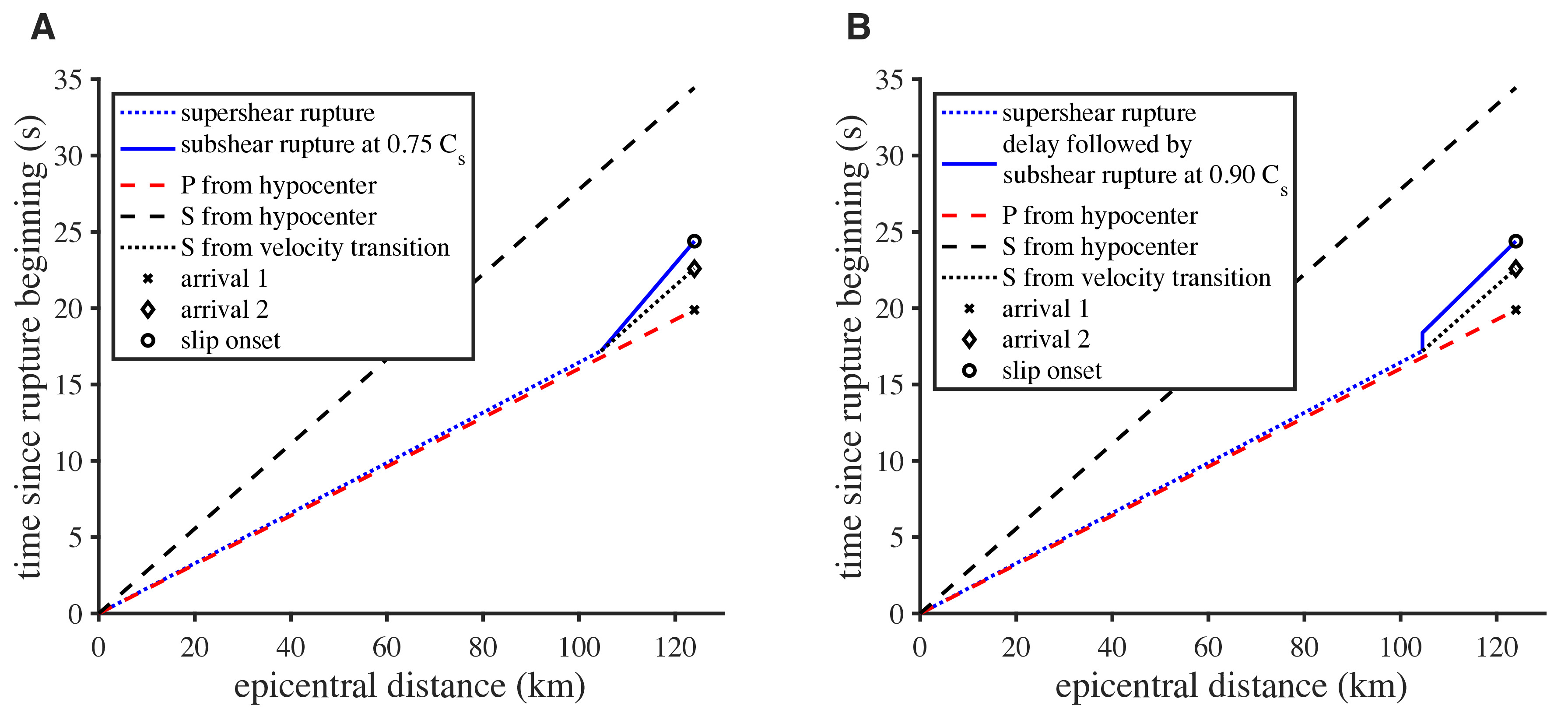}
	\caption{\textbf{Two possible rupture histories of rupture propagation between the epicenter and the camera location, built to constrain the dynamic pulse model.} \textbf{(A)} Partially supershear model with a first supershear segment of 104.6 km at 0.975 $C_p$, and a second subshear segment at $0.75C_s$ fully subshear model at 0.9 $C_s$. \textbf{(B)} Same initial scenario as \textbf{(A)} with a stronger deceleration of the rupture, allowing a faster rupture velocity at the camera location.}
	\label{fig:SP3} 
\end{figure}

We need a rupture velocity as an input for the slip-pulse model. We build two simple rupture history models (Figure \ref{fig:SP3}) that are consistent with the observed wave arrivals before the slip onset at the camera location and the rupture time  measured at station NPW.

The rupture is propagating southward along this segment of the fault. Although we did not directly observe the rupture front in the video, there appear to be two consecutive frames between which the rupture front advances southward. However, due to the high rupture velocity, the low frame rate, and the limited resolution of the footage, it is not possible to make a definitive observation, particularly given the emergent nature of the slip function. Moreover, the camera's internal clock is not synchronised with official timing: the timestamps embedded in the footage are approximately four minutes earlier than the event origin time reported by the USGS \cite{ref9}. This discrepancy prevents any direct estimation of the local or average rupture velocity from the video recording.

Kinematic models have proposed supershear rupture propagation along this segment of the fault~\cite{ref9,ref11}.The seismic station NPW, located $2$~km off-fault, $124$~km south of the site location (see figure \ref{fig:SP2}) and $248$~km south of the epicenter, registered the passage of the rupture front and constrains an average rupture velocity of $4.8$~km.s$^{-1}$ for the first $248$~km of rupture propagation to the south, which is supershear.

However, it is clear from the footage that the second observed arrival, occurring approximately $1.8$~s before the onset of slip, corresponds to a strong horizontal ground motion to the East, in a fault-normal direction. This strong fault normal motion is expected for sub-shear pulses on the fault plane, just ahead of the rupture front, in mode II steady state slip-pulse models ~\cite{ref49,ref50}. It corresponds to near-field deformation ahead of the rupture tip. 
Therefore this suggests that, at this location, the rupture front propagated at subshear velocity. Yet, dueto the high average rupture velocity between the epicenter and NPW ($\approx 4.8$~km/s), we cannot assume subshear velocity from the epicenter to the camera location. This would imply velocities larger than the P-wave speed between the camera and NPW, which is not physically possible. 

We thus propose a model in which there is a transition from supershear to subshear velocity before the rupture reaches the camera location. In this model (Figure \ref{fig:SP3}), the first arrival, at $t=-4.5$~s, is interpreted as the P-wave radiated from the hypocenter. The second arrival is interpreted as the S-wave radiated from the supershear-subshear transition. With $C_s=3600$~m/s and $C_p=6235$~m/s, several models are possible. This interpretation suggests a wide range of possible local rupture velocities at the camera location. To span the largest range, we assumed a very high rupture velocity of $0.975\,C_p$ during the first phase, which implies a sudden deceleration at approximately $19.6$~km north of the video location. Then, we propose two end-member scenarios. In the first one, the rupture velocity is constant after its deceleration, and can be as low as $V_r=0.75\,C_s$ (Figure \ref{fig:SP3}A). In the second one, a very strong deceleration of the rupture imposes a delay at the deceleration location, and then the rupture starts again with a new velocity. In this case, depending on the delay duration, the local velocity can be as high as the Rayleigh wave speed, but we chose a maximum value of $V_r=0.90\,C_s$ (Figure \ref{fig:SP3}B).

In all cases, the rupture must transition again to supershear velocity after its passage to the camera location, to remain coherent with the observed arrival time at NPW. More complex scenarios, involving two-dimensional rupture front geometries or more velocity changes, may also account for the observed arrival times, though such considerations are beyond the scope of this study.

\subsection{Elastodynamic pulse models}

\subsubsection*{Steady state sub-Rayleigh slip pulse with imposed slip-rate function}

In this approach, the observed slip-rate function is imposed as a boundary condition, and the resulting stress evolution is computed.

\subsubsection{Model}
We consider a 2D plane-strain (mode II) shear rupture propagating in a homogeneous, isotropic linear elastic medium characterized by a shear modulus $\mu = 35$~GPa, and Poisson's ratio $\nu = 0.25$. The rupture propagates at a constant sub-Rayleigh velocity $V_r$, as a pulse of size $L$. We focus on the southern rupture front, assuming that the northern front is located at a distance much greater than $L$, such that its influence can be neglected. The rupture is assumed to be in steady state, meaning that all physical quantities are stationary in a frame co-moving with the rupture tip at velocity $V_r$. In this co-moving frame, we define the spatial coordinate $x$ as the distance from the tip of the pulse, with $x>0$ corresponding to points inside the slipping region (i.e., behind the tip), and $x<0$ corresponding to points ahead of the rupture front.

In this framework, the shear stress $ \tau(x) $ along the fault is related to the slip rate distribution $ v(x) $ by the singular integral equation \cite{ref29,ref51}:
\begin{equation}
\label{eq:elastodynamics}
\tau(x) = \tau_0 -\frac{\bar{\mu}}{2\pi V_r} \, \mathrm{P.V.} \int_{0}^{L} \frac{v(\xi)}{x - \xi} \, d\xi,
\end{equation}
where $\tau_0$ is the background stress, $x = V_r (t-t_i)$ is the distance to the rupture tip, $t_i$ is the exact time of the slip onset, $L = V_r\Delta t$ is the length of the slip pulse, which defines the spatial extent of the actively slipping region, and $\mathrm{PV}$ denotes the Cauchy principal value of the integral. $\bar{\mu}$ is a scaled shear modulus:

\begin{equation}
    \bar{\mu} = \mu \frac{4\alpha_s\alpha_d - (1+\alpha_s^2)^2}{\alpha_s(1-\alpha_s^2)}
\end{equation}
where $\alpha_s = \sqrt{1-(V_r/C_s)^2}$ and $\alpha_p = \sqrt{1-(V_r/C_p)^2}$. This expression represents the elastodynamic balance between dynamic stress and radiated fields generated by a spatially varying slip rate. The integral is weakly singular and captures the long-range elastic interactions typical of rupture problems.

To evaluate this expression numerically, we map the physical domain $ x \in [0, L] $ to the interval $ X  = 2x/L - 1 \in [-1, 1] $, and use a Gauss-Chebyshev quadrature \cite{ref52,ref30}, with Chebyshev polynomials of the second kind $U_n$. Using $N$ nodes $ S_j $ and weights $ w_j $, the discretized stress at evaluation points $ X_i $ becomes:
\begin{equation}
\label{eq:discrete_elastodynamics}
\Delta\tau(X_i) =  \tau(X_i) - \tau_0 = -\frac{\bar{\mu}}{2\pi V_r} \sum_{j=1}^{N} \frac{w_j}{X_i - S_j} \cdot \frac{v(S_j)}{\sqrt{1 - S_j^2}},
\end{equation}
where $ w_j $, $ S_j $, and $ X_i $ are given in Table~1 of \cite{ref52}.

The slip rate values $v(S_j)$ at the quadrature points $s_j$ are obtained by interpolating the slip rate measurements $V_k$ of Figure~\ref{fig:3}B, recorded at discrete times $t_k \in [t_i, t_f]$, with $t_i = 0.233$~s and $t_f=t_i + \Delta t$, where the pulse duration is $\Delta t = 1.4$~s. To this end, we define the reduced positions as $X_k = 2V_r t_k / L - 1$, and fit the data using linear least squares on the basis of functions $\phi_p(X) = \sqrt{1 - X^2} , U_p(X)$, where $p \in [0, P]$. This particular choice of basis ensures that the inferred shear stress remains finite at both the leading and trailing edges of the slip pulse. To mitigate artifacts introduced by signal smoothing, the interpolation is fitted to the cumulative slip data $D_k = \delta(t_k)$ (Figure~\ref{fig:4}B), rather than directly to the slip rate $V_k = v(t_k)$. In addition, continuity of the stress gradient is enforced at the trailing edge of the pulse ($x = L$), as prescribed by fault mechanics \cite{ref51}.
Moreover, we impose continuity of the stress gradient at the trailing edge of the pulse $x=L$ \cite{ref51}. The code \texttt{figure\_4\_S5\_S6\_S7.py} developed to perform the elastodynamic inversion are available online \cite{ref53}.

\subsubsection{Sensitivity analysis}

The inversion depends primarily on five parameters: (1) the maximum polynomial degree $P$ of the Chebyshev-based interpolation basis, (2) the arrival time $t_i$ of the pulse, (3) the pulse duration $\Delta t$, (4) the normalized rupture velocity $V_r/C_s$, and (5) uncertainties in fault slip estimation from image analysis. We systematically assess the influence of each parameter on the inferred source properties, namely the slip-weakening distance $D_c$, the pulse length $L$, the cohesive zone size $X_c$, and the breakdown work $W_b$.

\begin{figure}
    \centering
    \includegraphics[width=\linewidth]{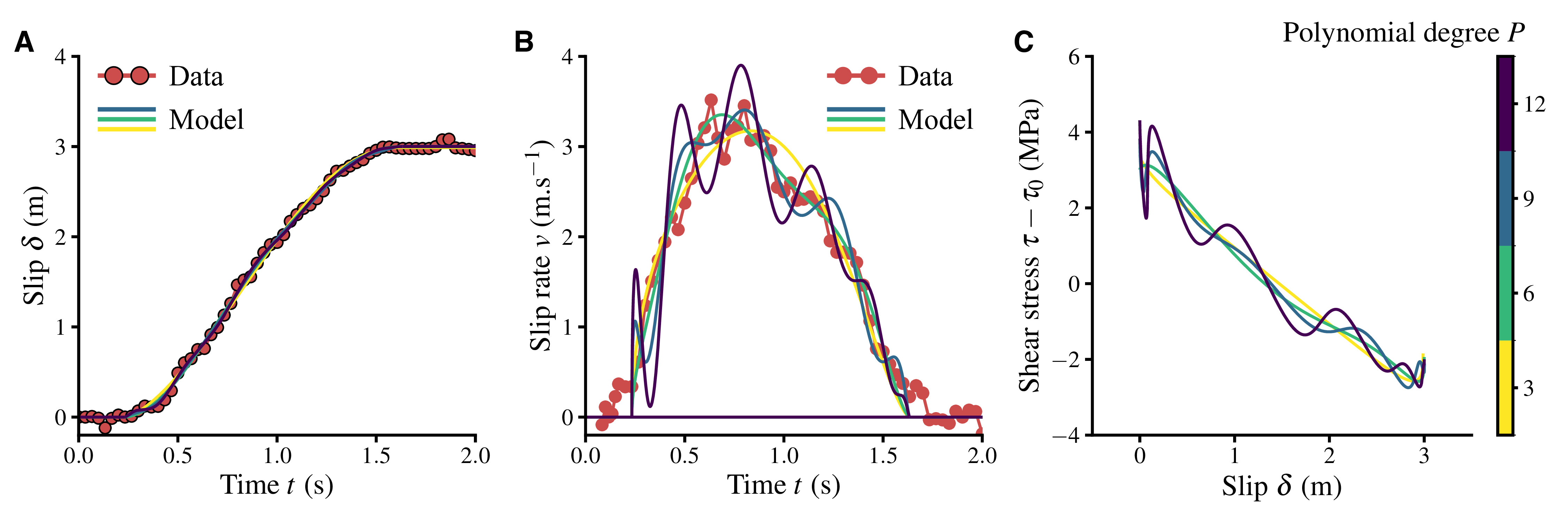}
    \caption{\textbf{(A)} Best-fit reconstruction (colored solid lines) of the cumulative slip $D_k = \delta(t_k)$ at discrete times (red circles, shown every third data point), using the basis $(\phi_p)_{p\in[0,P]}$ with increasing polynomial degree $P$. Fitting is performed using $t_i = 0.233$~s, $\Delta t = 1.4$~s, and $V_r = 0.90\,C_s$. \textbf{(B)} Interpolated slip rate (colored dashed lines) compared to the discrete-time measurements $V_k = v(t_k)$ (red circles, every third point). \textbf{(C)} Inferred shear stress change $\tau - \tau_0$ as a function of slip $\delta$.}
    \label{fig:SP4}
\end{figure}

First, we tested maximum polynomial degree $P \in [3, 12]$ in Figure~\ref{fig:SP4}, and found that their influence on the inferred shear stress $\tau(x)$ was minimal. We selected $P = 6$ as a balance between capturing the slip-rate asymmetry (absent at low $P$), and avoiding high-frequency oscillations that emerge at larger $P$ and are not supported by the data.

\begin{figure}
    \centering
    \includegraphics[width=\linewidth]{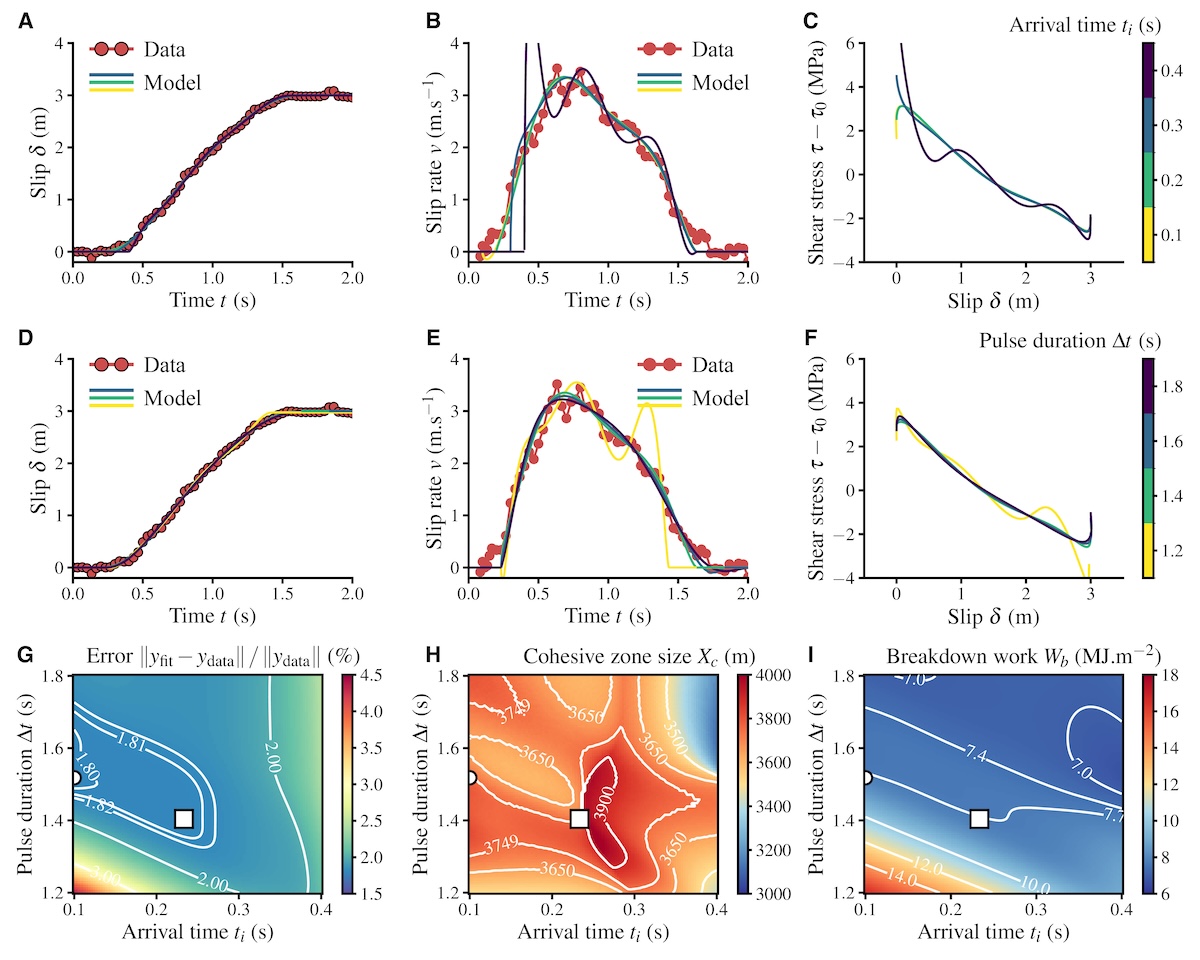}
    \caption{Best-fit reconstruction of cumulative slip $\delta$, interpolated slip rate $v$, and inferred shear stress change $\tau - \tau_0$, for varying \textbf{(A–C)} arrival time $t_i$ and \textbf{(D–F)} pulse duration $\Delta t$. Inversions are performed assuming $V_r = 0.90\,C_s$. Discrete measurements from image analysis are shown as red circles (every third data point), and elastodynamic inversion results are plotted as solid colored lines. \textbf{(G)} Error on slip measurements, \textbf{(H)} cohesive zone size $X_c$, and \textbf{(I)} breakdown work $W_b$ as functions of $t_i$ and $\Delta t$. The global minimum over $(t_i, \Delta t)$ is indicated by a white circle; parameters used in the main text are marked with a white square. Main text reports $X_c = 3749$~m and $W_b = 7.7$~MJ/m$^2$.}
    \label{fig:SP5}
\end{figure}

Next, we assess the sensitivity of our results to the arrival time $t_i$ and pulse duration $\Delta t$. In the image-based measurements, slip onset is first detected from the motion of poles behind the fence (Figure~\ref{fig:2}A) at $t_i = 0.233$~s. No discernible displacement is observed after $t_f = 1.666 \pm 0.067$~s, yielding an estimated pulse duration of $\Delta t \simeq 1.4$~s. Theoretical scaling laws suggest that slip evolves as $\delta(t) \propto (t - t_i)^{3/2}$ at the leading edge and $\delta(t) \propto (t_f - t)^{5/2}$ at the trailing edge \cite{ref51}, implying that the temporal and spatial resolution of the CCTV footage may introduce uncertainty in the precise determination of $t_i$ and $\Delta t$, and ultimately on the source parameters. To quantify this uncertainty, we tested a range of values $t_i \in [0.1, 0.4]$~s and $\Delta t \in [1.2, 1.8]$~s (Figure\ref{fig:SP5}). We find that, except at the extremal values of these ranges, most $(t_i, \Delta t)$ pairs yield good fits to the observed slip data (Figure\ref{fig:SP5}G), with minimal impact on the inferred $D_c$ (Figures~\ref{fig:SP5}A–F), $X_c$ (Figure~\ref{fig:SP5}H), and $W_b$ (Figure~\ref{fig:SP5}I). Only the pulse length $L = V_r \Delta t$ remains directly dependent on the choice of $\Delta t$. In the absence of more precise constraints, we adopt the image-derived value $t_i = 0.233$~s, for which the pulse duration $\Delta t = 1.4$~s provides the best fit to the observed slip, consistent with the image-based estimate. While our measurement of $L$ is sensitive to the assumed arrival time $t_i$ and pulse duration $\Delta t$, estimates of slip-weakening distance $D_c$ and cohesive zone size $X_c$ remains robust across a wide range of $(t_i, \Delta t)$ values (Figure~\ref{fig:SP5}). This makes $D_c$ and $X_c$ reliable parameters for calibrating frictional constitutive laws using dynamic rupture simulations.

\begin{figure}
    \centering
    \includegraphics[width=\linewidth]{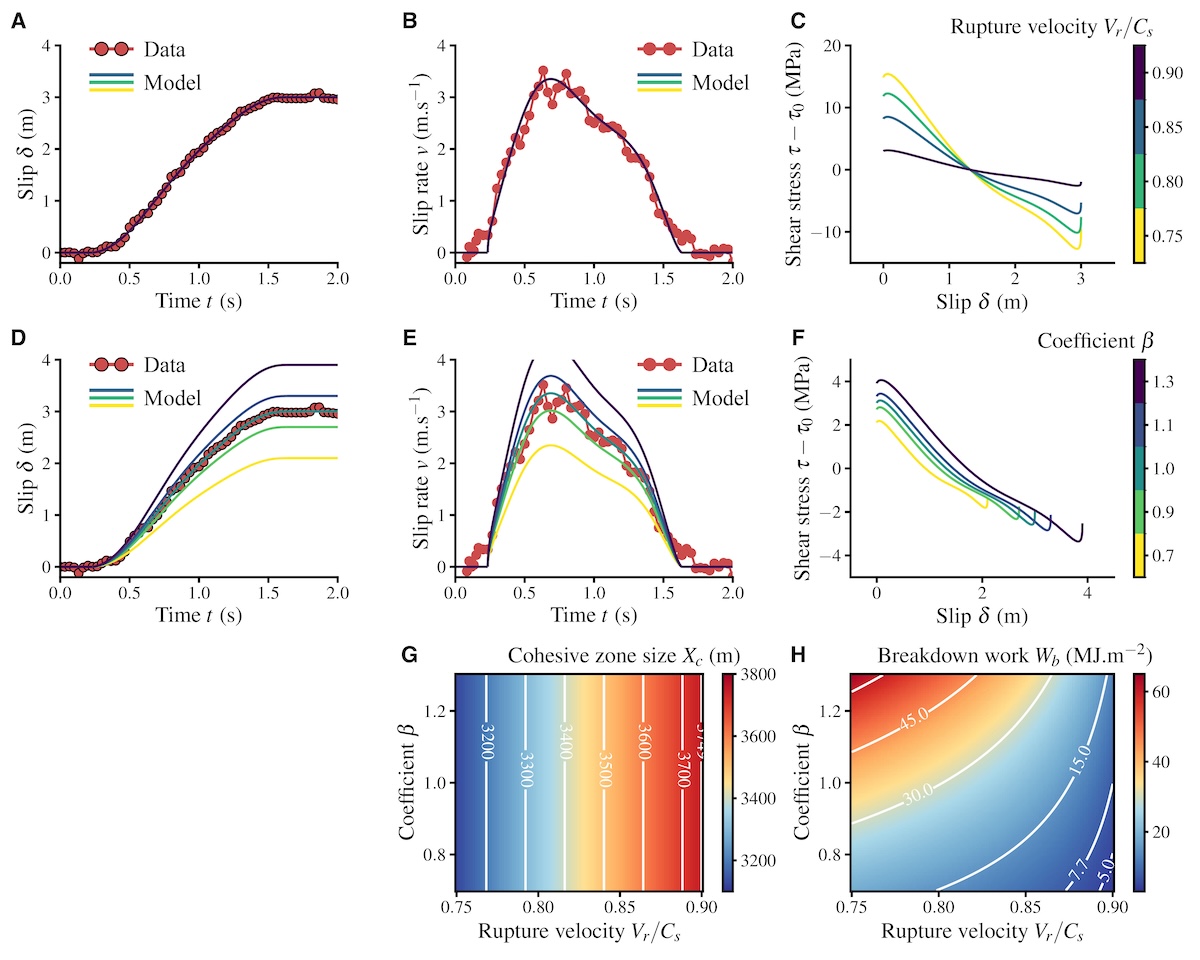}
    \caption{Best-fit reconstruction of cumulative slip $\delta$, interpolated slip rate $v$, and inferred shear stress change $\tau - \tau_0$, for varying \textbf{(A–C)} scaled rupture velocity $V_r/C_s$ and \textbf{(D–F)} slip scaling factor $\beta$. Inversions are performed assuming $t_i = 0.233$~s and $\Delta t=1.4$~s. Discrete measurements from image analysis are shown as red circles (every third data point), and elastodynamic inversion results are plotted as solid colored lines. \textbf{(G)} Cohesive zone size $X_c$, and \textbf{(H)} breakdown work $W_b$ as functions of $V_r/C_s$ and $\beta$. Main text reports $X_c = 3749$~m and $W_b = 7.7$~MJ/m$^2$.}
    \label{fig:SP6}
\end{figure}

Finally, we evaluate how uncertainties in slip rate and rupture velocity affect the estimation of breakdown work $W_b$. As shown in Figure~\ref{fig:SP6}, we explore a range of scaled rupture velocities $V_r/C_s \in [0.75, 0.90]$, and vary the final slip using a scaling factor $\beta \in [0.7, 1.3]$, consistent with the identified rupture scenarios and the uncertainty bounds shown in Figure~\ref{fig:3}. As expected, the slip-weakening distance $D_c$ scales linearly with $\beta$, but remains insensitive to variations in $V_r$. In contrast, the cohesive zone size $X_c$ increases linearly with $V_r$, due to the normalization of spatial dimensions by the pulse length $L = V_r \Delta t$, but is unaffected by changes in $\beta$. The breakdown work $W_b$ is highly sensitive to both $V_r/C_s$ and $\beta$, with estimated values ranging from $3.8$~MJ/m$^2$ to $64.5$~MJ/m$^2$.

\subsection{Steady state sub-Rayleigh slip pulse with imposed friction law}

In this approach, the stress distribution corresponding to a classical friction law is imposed on the fault as a boundary condition, and the resulting particle velocity field is computed. Key input parameters are inverted so that this velocity field fit to the observed slip rate function.

\subsubsection{Model}
We consider a 2D slip pulse of length $L$ propagating at a constant rupture velocity, $v_r$, in a homogeneous, isotropic linear elastic medium characterized by a shear modulus, $\mu$, and a Poisson's ratio, $\nu$ (Figure S3). The slip pulse lies on the $y=0$ plane. We assume that the frictional strength decays linearly from a peak value, $\tau_p$, to a residual value, $\tau_r$ over a cohesive zone, $X_c$ (Figure \ref{fig:SP7}). Following the model proposed by Rice and his collaborators \cite{ref29} we can calculate the particle velocities induced by this slip pulse, at a location $(x,y)$ as

\begin{align}
  v_x(x,y) = -\dfrac{(\tau_p-\tau_r)v_r}{\mu D}\left\{2\alpha_s \mathrm{Im}[M(z_p,X_c,L)] - \alpha_s(1+\alpha_s^2)\mathrm{Im}[M(z_s,X_c,L)]\right\}\\
v_y(x,y) = -\dfrac{(\tau_p-\tau_r)v_r}{\mu D}\left\{2\alpha_s \alpha_p\mathrm{Re}[M(z_p,X_c,L)] - (1+\alpha_s^2)\mathrm{Re}[M(zs,X_c,L)]\right\}   
\end{align}

Here $\alpha_s$ and $\alpha_p$ are defined earlier, $D$ is the Rayleigh function defined as $[4\alpha_s\alpha_d - (1+\alpha_s^2)^2]$. $\mathrm{Re}[M(z)]$ and $\mathrm{Im}[M(z)]$ correspond to the real and the imaginary parts of the analytic function $M(z)$. $z_p = x+i\alpha_p y$ and $z_s = x+i\alpha_s y$.

The analytic function $M(z)$ is given by,

\begin{equation}
    M(z) = -\frac{1}{\pi}\sqrt{z(z+L)}\int_{-L}^0 { \frac{\tau_f(\xi)}{\sqrt{-\xi(\xi+L)}(\xi-z)} d\xi }
\end{equation}

Here $\tau_f(\xi)$ is the cohesive zone law prescribed behind the rupture front. When $\tau_{f}(\xi)$ is linear, we can obtain the closed-form expression for $M(z)$ and hence the complete solution. The slip rate on the fault can now easily be estimated as $v(x) = 2v_x(x,y\to0^+)$, from which the slip, $\delta(x)$, can be calculated, which can be used to compute an equivalent slip-weakening friction law.

From basic energy balance of a slip-pulse, we can obtain the energy release rate, $G$, as $G = \int_0^{D_c} [\tau_f(\delta) - \tau_r]d\delta$ .

\subsubsection{Generalization to a supershear case}
Because the ground motions recorded by the camera indicates that the rupture likely propagated at sub-Rayleigh speed at the observation site, we do not include rupture properties for supershear scenarios in the main text. However, our second model is capable of simulating rupture at supershear velocities. In the Supplementary Material, for the sake of completeness, we present a simulation assuming a constant supershear rupture velocity between the epicenter and station NPW, located 248 km away. Assuming a rupture arrival time of 51 seconds at NPW, the corresponding average rupture velocity is approximately $1.3\,C_s$.

\begin{figure}
    \centering
    \includegraphics[width=1\textwidth]{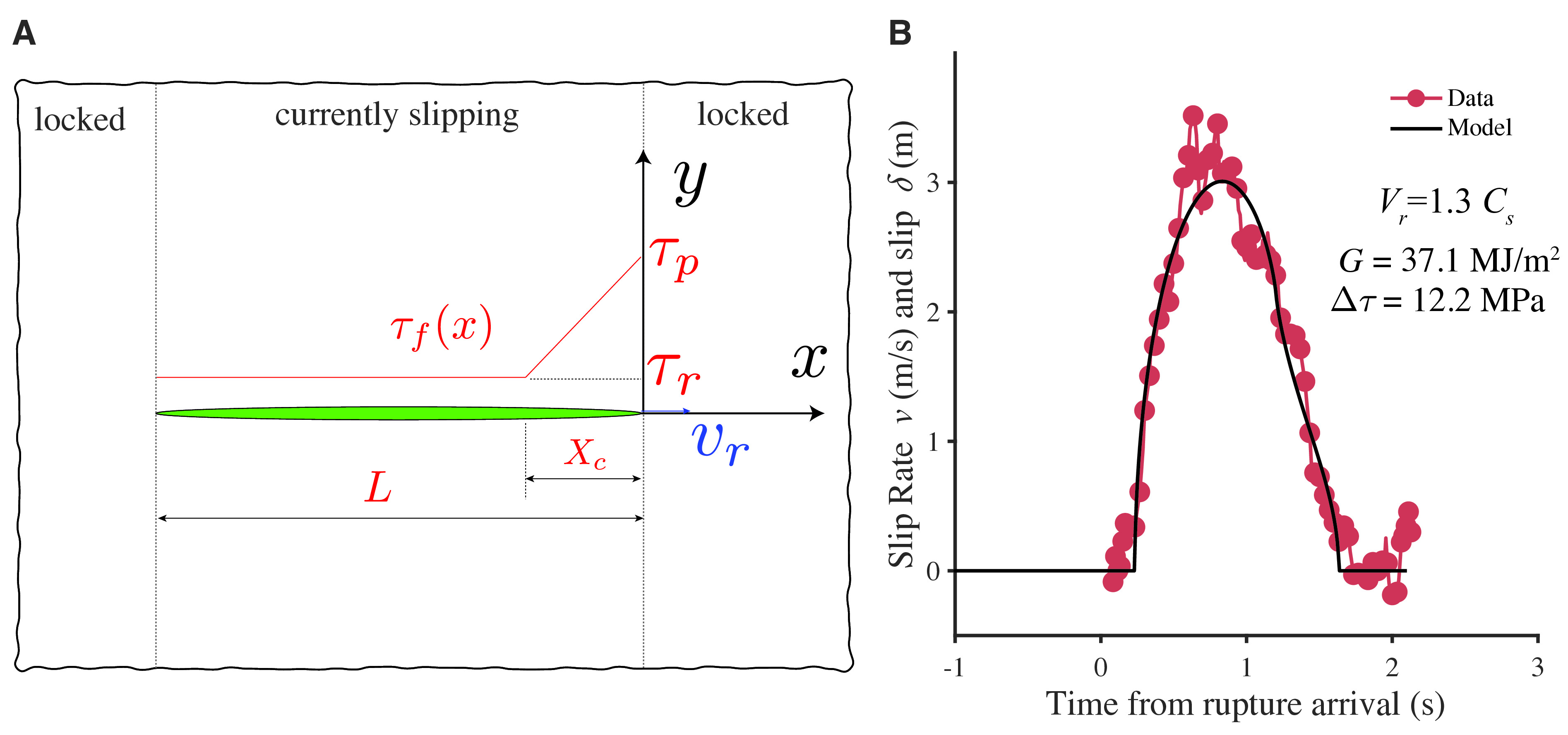}
    \caption{Steady state slip pulse model adapted from Rice and collaborators \cite{ref29}. \textbf{(A)} Scheme of the rupture pulse model. \textbf{(B)} Best fit obtained assuming the supershear rupture scenario, and associated rupture properties.}
    \label{fig:SP7}
\end{figure}

Using this velocity and applying the same cohesive law as in the sub-Rayleigh case, the best fit to the observed data is achieved with a background normal stress of $\sigma_{yy}^{0} = 60$ MPa and a scaled cohesive zone size of $X_c/L = 0.71$ (Figure~\ref{fig:SP7}B). The inferred strength drop is $\tau_p - \tau_r = 30$ MPa, the stress drop is $\sigma_{xy}^{0} - \tau_r = 12.2$ MPa, and the resulting slip-weakening distance is $D_c \approx 2.57$ m. The energy release rate, computed as $G = \int_0^{D_c} [\tau_f(\delta) - \tau_r],d\delta$, is found to be $G = 37.1$ MJ/m$^2$.

\clearpage
\renewcommand{\refname}{References and Notes}


\begin{thebibliography}{99}

\bibitem{ref1} J.-P. Avouac, From geodetic imaging of seismic and aseismic fault slip to dynamic modeling of the seismic cycle. \textit{Annual Review of Earth and Planetary Sciences} \textbf{43} (1), 233--271 (2015).

\bibitem{ref2} A. Tarantola, B. Valette, Generalized nonlinear inverse problems solved using the least squares criterion. \textit{Reviews of Geophysics} \textbf{20} (2), 219--232 (1982).

\bibitem{ref3} A. Sarao, S. Das, P. Suhadolc, Effect of non-uniform station coverage on the inversion for earthquake rupture history for a Haskell-type source model. \textit{Journal of Seismology} \textbf{2}, 1--25 (1998).

\bibitem{ref4} I. A. Beresnev, Uncertainties in finite-fault slip inversions: to what extent to believe? (a critical review). \textit{Bulletin of the Seismological Society of America} \textbf{93} (6), 2445--2458 (2003).

\bibitem{ref5} S. Hartzell, P. Liu, C. Mendoza, C. Ji, K. M. Larson, Stability and uncertainty of finite-fault slip inversions: Application to the 2004 Parkfield, California, earthquake. \textit{Bulletin of the Seismological Society of America} \textbf{97} (6), 1911--1934 (2007).

\bibitem{ref6} P. M. Mai, et al., The earthquake-source inversion validation (SIV) project. \textit{Seismological Research Letters} \textbf{87} (3), 690--708 (2016).

\bibitem{ref7} G. Bertrand, C. Rangin, Tectonics of the western margin of the Shan plateau (central Myanmar): implication for the India--Indochina oblique convergence since the Oligocene. \textit{Journal of Asian Earth Sciences} \textbf{21} (10), 1139--1157 (2003).

\bibitem{ref8} N. G. Reitman, et al., ArcGIS Experience Builder -- Web App (2025), \url{https://experience.arcgis.com/experience/e40a6967c3ea42dd85bf44037e05482b}, consulted on May 15, 2025.

\bibitem{ref9} U.S. Geological Survey, Finite Fault -- M 7.7 - 2025 Mandalay, Burma (Myanmar) Earthquake (2025), \url{https://earthquake.usgs.gov/earthquakes/eventpage/us7000pn9s/finite-fault}, consulted on May 16, 2025.

\bibitem{ref10} M. Vallee, J. Charléty, A. M. Ferreira, B. Delouis, J. Vergoz, SCARDEC: a new technique for the rapid determination of seismic moment magnitude, focal mechanism and source time functions for large earthquakes using body-wave deconvolution. \textit{Geophysical Journal International} \textbf{184} (1), 338--358 (2011).

\bibitem{ref11} N. Inoue, et al., A multiple asymmetric bilateral rupture sequence derived from the peculiar tele-seismic P-waves of the 2025 Myanmar earthquake (2025).

\bibitem{ref12} H. Richter, Watch an earthquake split a hillside in two. \textit{Science} (2025), news article published on the Science website, doi:10.1126/science.zt83qxx.

\bibitem{ref13} J. Kearse, Y. Kaneko, Curved Fault Slip Captured by CCTV Video During the 2025 Mw 7.7 Myanmar Earthquake. \textit{The Seismic Record} \textbf{5} (3), 281--288 (2025).

\bibitem{ref14} T. H. Heaton, Evidence for and implications of self-healing pulses of slip in earthquake rupture. \textit{Physics of the Earth and Planetary Interiors} \textbf{64}, 1--20 (1990).

\bibitem{ref15} D. Melgar, G. P. Hayes, Systematic Observations of the Slip Pulse Properties of Large Earthquake Ruptures. \textit{Geophysical Research Letters} \textbf{44} (19), 9691--9698 (2017), doi:10.1002/2017GL074916.

\bibitem{ref16} V. Lambert, N. Lapusta, S. Perry, Propagation of large earthquakes as self-healing pulses or mild cracks. \textit{Nature} \textbf{591} (7849), 252--258 (2021).

\bibitem{ref17} F. Barras, E. Aharonov, F. Renard, A minimal model illuminates the physics of pulse-like seismic rupture and oscillatory slip rates in damaged faults. \textit{Geophysical Research Letters} \textbf{52} (4), e2024GL111189 (2025).

\bibitem{ref18} H. Weng, J.-P. Ampuero, The dynamics of elongated earthquake ruptures. \textit{Journal of Geophysical Research: Solid Earth} \textbf{124} (8), 8584--8610 (2019).

\bibitem{ref19} H. Weng, J.-P. Ampuero, Integrated rupture mechanics for slow slip events and earthquakes. \textit{Nature Communications} \textbf{13} (1), 7327 (2022).

\bibitem{ref20} J. R. Rice, Heating and weakening of faults during earthquake slip. \textit{Journal of Geophysical Research: Solid Earth} \textbf{111} (B5) (2006).

\bibitem{ref21} A.-A. Gabriel, D. I. Garagash, K. H. Palgunadi, P. M. Mai, Fault size--dependent fracture energy explains multiscale seismicity and cascading earthquakes. \textit{Science} \textbf{385}(6707), eadj9587 (2024).

\bibitem{ref22} D. Garagash, Seismic and aseismic slip pulses driven by thermal pressurization of pore fluid. \textit{Journal of Geophysical Research: Solid Earth} \textbf{117} (B4) (2012).

\bibitem{ref23} S. Minson, M. Simons, J. Beck, Bayesian inversion for finite fault earthquake source models I Theory and algorithm. \textit{Geophysical Journal International} \textbf{194} (3), 1701--1726 (2013).

\bibitem{ref24} H. N. Razafindrakoto, P. M. Mai, Uncertainty in earthquake source imaging due to variations in source time function and earth structure. \textit{Bulletin of the Seismological Society of America} \textbf{104} (2), 855--874 (2014).

\bibitem{ref25} I. A. Beresnev, Choices of slip function and simulated ground motions. \textit{Pure and Applied Geophysics} \textbf{181} (6), 1859--1869 (2024).

\bibitem{ref26} G. Lykotrafitis, A. J. Rosakis, G. Ravichandran, Self-healing pulse-like shear ruptures in the laboratory. \textit{Science} \textbf{313} (5794), 1765--1768 (2006).

\bibitem{ref27} S. Nielsen, R. Madariaga, On the Self-Healing Fracture Mode. \textit{Bulletin of the Seismological Society of America} \textbf{93} (6), 2375--2388 (2003).

\bibitem{ref28} E. Tinti, E. Fukuyama, A. Piatanesi, M. Cocco, A Kinematic Source-Time Function Compatible with Earthquake Dynamics. \textit{Bulletin of the Seismological Society of America} \textbf{95}(4), 1211--1223 (2005).

\bibitem{ref29} J. R. Rice, C. G. Sammis, R. Parsons, Off-Fault Secondary Failure Induced by a Dynamic Slip Pulse. \textit{Bulletin of the Seismological Society of America} \textbf{95} (1), 109--134 (2005), doi:10.1785/0120030166.

\bibitem{ref30} N. Brantut, D. I. Garagash, H. Noda, Stability of pulse-like earthquake ruptures. \textit{Journal of Geophysical Research: Solid Earth} \textbf{124} (8), 8998--9020 (2019).

\bibitem{ref31} E. M. Dunham, R. J. Archuleta, Evidence for a supershear transient during the 2002 Denali fault earthquake. \textit{Bulletin of the Seismological Society of America} \textbf{94} (6B), S256--S268 (2004).

\bibitem{ref32} F. Paglialunga, F. Passelegue, M. Lebihain, M. Violay, Frictional weakening leads to unconventional singularities during dynamic rupture propagation. \textit{Earth and Planetary Science Letters} \textbf{626}, 118550 (2024).

\bibitem{ref33} D. S. Kammer, et al., Earthquake energy dissipation in a fracture mechanics framework. \textit{Nature communications} \textbf{15} (1), 4736 (2024).

\bibitem{ref34} B. Fryer, M. Lebihain, C. Noel, F. Paglialunga, F. Passelègue, The effect of stress barriers on unconventional-singularity-driven frictional rupture. \textit{Journal of the Mechanics and Physics of Solids} \textbf{193}, 105876 (2024).

\bibitem{ref35} R.E. Abercrombie, J.R. Rice, Can observations of earthquake scaling constrain slip weakening? \textit{Geophysical Journal International} \textbf{162} (2), 406--424 (2005).

\bibitem{ref36} E. Tinti, P. Spudich, M. Cocco, Earthquake fracture energy inferred from kinematic rupture models on extended faults. \textit{Journal of Geophysical Research: Solid Earth} \textbf{110} (B12) (2005).

\bibitem{ref37} S. Nielsen, et al., G: Fracture energy, friction and dissipation in earthquakes. \textit{J. Seismol.} \textbf{20}, 1187--1205 (2016), doi:10.1007/s10950-016-9560-1.

\bibitem{ref38} M. Cocco, et al., Fracture energy and breakdown work during earthquakes. \textit{Annual Review of Earth and Planetary Sciences} \textbf{51} (1), 217--252 (2023).

\bibitem{ref39} I. Svetlizky, J. Fineberg, Classical shear cracks drive the onset of dry frictional motion. \textit{Nature} \textbf{509} (7499), 205--208 (2014).

\bibitem{ref40} F. Paglialunga, et al., On the scale dependence in the dynamics of frictional rupture: Constant fracture energy versus size-dependent breakdown work. \textit{Earth and Planetary Science Letters} \textbf{584}, 117442 (2022).

\bibitem{ref41} D. S. Kammer, G. C. McLaskey, Fracture energy estimates from large-scale laboratory earthquakes. \textit{Earth and Planetary Science Letters} \textbf{511}, 36--43 (2019).

\bibitem{ref42} S. Xu, E. Fukuyama, F. Yamashita, Robust estimation of rupture properties at propagating front of laboratory earthquakes. \textit{Journal of Geophysical Research: Solid Earth} \textbf{124} (1), 766--787 (2019).

\bibitem{ref43} A. Tsutsumi, T. Shimamoto, High-velocity frictional properties of gabbro. \textit{Geophysical Research Letters} \textbf{24} (6), 699--702 (1997).

\bibitem{ref44} T. Hirose, T. Shimamoto, Growth of molten zone as a mechanism of slip weakening of simulated faults in gabbro during frictional melting. \textit{Journal of Geophysical Research: Solid Earth} \textbf{110} (B5) (2005).

\bibitem{ref45} G. Di Toro, et al., Fault lubrication during earthquakes. \textit{Nature} \textbf{471} (7339), 494--498 (2011).

\bibitem{ref46} T. Mikumo, K. B. Olsen, E. Fukuyama, Y. Yagi, Stress-breakdown time and slip-weakening distance inferred from slip-velocity functions on earthquake faults. \textit{Bulletin of the Seismological Society of America} \textbf{93} (1), 264--282 (2003).

\bibitem{ref47} Department of Meteorology and Hydrology - National Earthquake Data Center, Myanmar National Seismic Network [Data set]. International Federation of Digital Seismograph Networks (2016), doi:10.7914/SN/MM.

\bibitem{ref48} GEOFON Data Centre, GEOFON Seismic Network [Data set]. GFZ Data Services (1993), doi:10.14470/TR560404.

\bibitem{ref49} E. M. Dunham, R. J. Archuleta, Near-source ground motion from steady state dynamic rupture pulses. \textit{Geophysical Research Letters} \textbf{32} (3) (2005).

\bibitem{ref50} T. Gabrieli, Y. Tal, Lab earthquakes reveal a wide range of rupture behaviors controlled by fault bends. Proceedings of the National Academy of Sciences \textbf{122} (17), e2425471122 (2025).

\bibitem{ref51} D. I. Garagash, L. N. Germanovich, Nucleation and arrest of dynamic slip on a pressurized fault. \textit{Journal of Geophysical Research: Solid Earth} \textbf{117} (B10) (2012).

\bibitem{ref52} R. C. Viesca, D. I. Garagash, Numerical methods for coupled fracture problems. \textit{Journal of the Mechanics and Physics of Solids} \textbf{113}, 13--34 (2018).

\bibitem{ref53} Passelègue, F., Data and model for Direct Estimation of Earthquake Source Properties from a Single CCTV Camera. Zenodo. doi:10.5281/zenodo.16785672 (2025).

\bibitem{ref54} Hudnut, K. Sieh, Behavior of the Superstition Hills fault during the past 330 years. \textit{Bulletin of the Seismological Society of America} \textbf{79} (2), 304--329 (1989)

\end{thebibliography}
\end{document}